# EMERGING METHODS AND TOOLS FOR SPARKING NEW GLOBAL CREATIVE NETWORKS

Jeff Horon

Elsevier Research Intelligence
360 Park Ave S
New York, NY, 10010, USA
e-mail: j.horon@elsevier.com

## ABSTRACT

Emerging methods and tools are changing the ways participants in global creative networks become aware of each other and proceed to interact. These methods and tools are beginning to influence the collaboration opportunities available to network participants.

Some web-based resources intended to spark new collaborations in creative networks have been plagued by dependence on fragmented or out-of-date information, having shallow recall (e.g. by being limited to a list of manually curated keywords), offering poor interconnectivity with other systems, and/or obtaining low end-user adoption.

Increased availability of information about creative network participants' activities and outputs (such as completed sponsored research projects and published results, aggregated into global databases), coupled with advancement in information processing techniques like Natural Language Processing (NLP), enables new web-based technologies for discovering subject matter experts, facilities, and networks of current and potential collaborators. Large-scale data resources and NLP allow modern versions of these tools to avoid the problems of having sparse/fragmented data and also provide for deep recall, sometimes within and across many disciplinary vocabularies. These tools are known as "passive" technologies, from the perspective of the creative network participant, because the agent must undertake an action to use the information resources placed at his or her disposal.

Emerging "active" methods and tools utilize the same types of information and technologies, but actively intervene in the formation of the creative network by suggesting connections and arranging virtual or physical interactions. Active approaches can achieve very high end-user adoption rates.

Both active and passive methods strive to use data-driven approaches to form better-than-chance awareness among networks of potential collaborators. Modern instances of both types of systems generally support interconnectivity with other systems, and therefore expand the size of participants' networks, resulting in a larger pool of potential collaborators from which to draw upon, within the system and additionally wherever the data is repurposed (e.g. into federated searches and customized applications).

## EXAMPLES AND CASE STUDIES

### "Passive" Networking Applications

The most widely deployed applications (providers) are: the Pure Experts portal (Elsevier), VIVO (DuraSpace), and Harvard Profiles (Harvard Medical School). Each of these applications facilitates search and discovery of subject matter experts and their research activities and outputs. These systems are generally organized and supported at the university level. These applications are also federated into multi-institutional search frameworks including Direct2Experts and CTSAsearch – both of which are open to all three of the networking applications above, as well as other less widely deployed applications.

### "Active" Networking Applications

Efforts toward active networking interventions are sometimes made with 'researcher speed dating' activities, but these generally rely on an audience with some mutual interests being gathered together (e.g. at a conference or symposium) and pairings are typically random. Despite the existence of predictive factors for propensity to collaborate and likelihood of achieving team goals (e.g. obtaining external funding for research projects)[i], data-driven active networking methods are comparatively rarely used. Prior case studies in active networking include:

*Team design for large center and team science proposals*

The University of Michigan Medical School assisted a principal investigator applicant for a large center grant with team formation, based on identifying potential participants publishing or having sponsored projects in subject matter related to the center. This allowed for discovery of related expertise by analyzing term co-occurrence, and then discovery of the subject matter experts working with those concepts. Multiple rounds of iteration resulted in a list of keywords, stemmed to related key terms, such that the list was both inclusive of the desired family of concepts and exclusive of 'false positive' matches.

*Suggested casual interactions at a physical event*

At an institute launch event, the University of Michigan employed search methods similar to those above for objective detection of researchers working in related topic areas, to supplement institute founders' knowledge of researchers working in relevant topic areas with information about previously-unknown researchers also working in these topic areas. Objective detection allowed for increased inclusiveness and comprehensiveness of the launch conference invitee list.

Launch event organizers solicited survey responses from participants concerning areas of methodological expertise, methodological needs for upcoming projects, and areas of interest within several pre-identified areas related to the institute.

Attendees were matched based upon expressing strong mutual interest in a topic and/or by study method, in situations where one researcher expressed a need for expertise in a method and another research expressed the ability to share methodological expertise in the same method. Reciprocal methodological need/provision matches were considered especially strong matches (Figure 1):

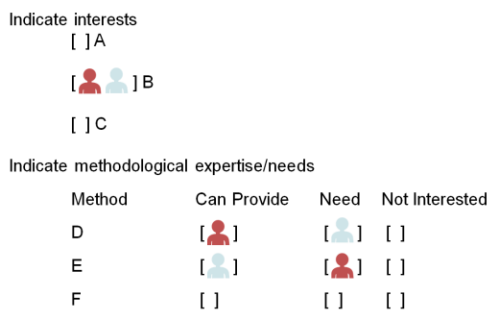

Figure 1: A generalized example of an especially strong match

Existing collaboration data covering co-authored publications and co-participation on sponsored projects were used to rule out matches who had collaborated in the past.

To maximize the chances strong matches would interact, the seating chart was also arranged to place strong matches at the same tables. This event also included conversation-provoking material, including a visualization of attendees arranged in a social networking diagram according to indicated areas of strong interest.

The matching process proved to be very flexible and was used to support a novel approach to bridging mentorship gaps in pediatric research[ii].

*Scheduled interactions at a physical event*

The University of Texas System M.D. Anderson Cancer Center has in recent years built into a key global cancer conference activities for scheduled networking interactions. The survey mechanism is similar to the University of Michigan example above, as are the recommendations, but there is also accommodation for arranging meetings including generally a mix of online meeting coordination, dedicated meeting time available, and dedicated meeting spaces available. Given rotating global locations and varied attendees from year-to-year, priority is given to matches from different institutions as there may only be one time they are physically co-located.

In addition to the meetings booked during a specific speed dating event window in the conference program, the project team also noted a number of off-hours and informal meetings taking place, driven in part by the recommended matches.

## **CONCLUSION**

These emerging methods and tools suggest the existence of repeatable strategies for facilitating data-driven matching and better-than-chance interactions designed to spark new global creative networks. As these methods become further systematized and see wider adoption, they are poised to influence larger numbers of creative networks and their participants.

---

[i] Lungeanu, A., Huang, Y., and Contractor, N.S. (2014) "Understanding the assembly of interdisciplinary teams and its impact on performance." *Journal of Informetrics*. **8(1)**:59-70.

[ii] Nigrovic, P.A., Muscal, E., Riebschleger, M., et. al. (2014) "AMIGO: A Novel Approach to the Mentorship Gap in Pediatric Rheumatology" *Journal of Pediatrics* **164(2)**:226-7.e1-3.